\PassOptionsToPackage{unicode}{hyperref}
\PassOptionsToPackage{hyphens}{url}
\documentclass[
]{article}
\usepackage{amsmath,amssymb}
\usepackage{resizegather}
\usepackage{lmodern}
\usepackage{iftex}
\ifPDFTeX
  \usepackage[T1]{fontenc}
  \usepackage[utf8]{inputenc}
  \usepackage{textcomp} 
\else 
  \usepackage{unicode-math}
  \defaultfontfeatures{Scale=MatchLowercase}
  \defaultfontfeatures[\rmfamily]{Ligatures=TeX,Scale=1}
\fi
\IfFileExists{upquote.sty}{\usepackage{upquote}}{}
\IfFileExists{microtype.sty}{
  \usepackage[]{microtype}
  \UseMicrotypeSet[protrusion]{basicmath} 
}{}
\makeatletter
\@ifundefined{KOMAClassName}{
  \IfFileExists{parskip.sty}{%
    \usepackage{parskip}
  }{
    \setlength{\parindent}{0pt}
    \setlength{\parskip}{6pt plus 2pt minus 1pt}}
}{
  \KOMAoptions{parskip=half}}
\makeatother
\usepackage{xcolor}
\usepackage{longtable,booktabs,array}
\usepackage{multirow}
\usepackage{calc} 
\usepackage{etoolbox}
\makeatletter
\patchcmd\longtable{\par}{\if@noskipsec\mbox{}\fi\par}{}{}
\makeatother
\IfFileExists{footnotehyper.sty}{\usepackage{footnotehyper}}{\usepackage{footnote}}
\makesavenoteenv{longtable}
\usepackage{graphicx}
\makeatletter
\def\maxwidth{\ifdim\Gin@nat@width>\linewidth\linewidth\else\Gin@nat@width\fi}
\def\maxheight{\ifdim\Gin@nat@height>\textheight\textheight\else\Gin@nat@height\fi}
\makeatother
\setkeys{Gin}{width=\maxwidth,height=\maxheight,keepaspectratio}
\makeatletter
\def\fps@figure{htbp}
\makeatother
\ifLuaTeX
  \usepackage{selnolig}  
\fi
\IfFileExists{bookmark.sty}{\usepackage{bookmark}}{\usepackage{hyperref}}
\IfFileExists{xurl.sty}{\usepackage{xurl}}{} 
\hypersetup{
  pdftitle={BayesGmed: An R-package for Bayesian Causal Mediation Analysis},
  pdfauthor={Belay B Yimer1*, Mark Lunt1, Marcus Beasley2, \\ Gary J Macfarlane2, John McBeth1},
  hidelinks,
  pdfcreator={LaTeX via pandoc}}

\title{\textbf{BayesGmed}: An R-package for Bayesian Causal Mediation Analysis}
\author{Belay B Yimer\textsuperscript{1}\textsuperscript{*}, Mark Lunt\textsuperscript{1},\\
Marcus Beasley\textsuperscript{2}, Gary J Macfarlane\textsuperscript{2},
John McBeth\textsuperscript{1}}
\date{}

\begin{document}
\maketitle

\textsuperscript{1}Centre for Epidemiology Versus Arthritis, University
of Manchester, Manchester, United Kingdom

\textsuperscript{2}Aberdeen Centre for Arthritis and Musculoskeletal
Health (Epidemiology Group), University of Aberdeen, Aberdeen, United
Kingdom

Corresponding author: belaybirlie.yimer@manchester.ac.uk

\hypertarget{abstracts}{%
\section*{Abstracts}\label{abstracts}}
The past decade has seen an explosion of research in causal mediation analysis. However, most analytic tools developed so far rely on frequentist methods which may not be robust in the case of small sample sizes. In this paper, we propose a Bayesian approach for causal mediation analysis based on Bayesian g-formula. We created \textbf{BayesGmed}, an R-package for fitting Bayesian mediation models in R. The application of the methodology (and software tool) is demonstrated by a secondary analysis of data collected as part of the MUSICIAN study, a randomised controlled trial of remotely delivered cognitive behavioural therapy (tCBT) for people with chronic pain. We tested the hypothesis that the effect of tCBT would be mediated by improvements in active coping, passive coping, fear of movement and sleep problems. The analysis of MUSICIAN data shows that tCBT has better-improved patients' self-perceived change in health status compared to treatment as usual (TAU). The adjusted log-odds of tCBT compared to TAU range from 1.491 (0.452, 2.612) when adjusted for sleep problems to 2.264 (1.063, 3.610) when adjusted for fear of movement. Higher scores of fear of movement (log-odds, -0.141 (-0.245, -0.048)), passive coping (log-odds, -0.217 (-0.351, -0.104)), and sleep problem (log-odds, -0.179 (-0.291, -0.078)) leads to lower odds of a positive self-perceived change in health status. The result of \textbf{BayesGmed}, however, shows that none of the mediated effects are statistically significant. We compared \textbf{BayesGmed} with the \textbf{mediation} R package, and the results were comparable. Finally, our probabilistic sensitivity analysis using the \textbf{BayesGmed} tool shows that the direct and total effect of tCBT persists even for a large departure in the assumption of no unmeasured confounding.

\hypertarget{introduction}{%
\section{Introduction}\label{introduction}}

Studies in the health and behavioural sciences often aim to understand
whether and, if so, how an intervention causes an outcome. The
randomised controlled trial is considered the most rigorous method for
answering the "whether" question, but often the "how" part remains
unclear. Causal mediation analysis plays an important role in
understanding the mechanism by which an intervention produces changes in
the outcome. Understanding how an intervention works can be key for
further improvement and targeting of an intervention program.

There is a fast-growing methodological literature on causal mediation
analysis {[}\emph{1-7}{]}. One of the most important developments in
mediation analysis is the incorporation of the causal inference approach
or the potential outcomes framework (POF) to estimate causal mediation
effects. This has led to (i) the formulation of different estimands
(effect definitions) that have explicitly causal interpretations, (ii)
clarification of the assumptions required for such effects to be
estimated from observed data, (iii) a framework for conducting
sensitivity analyses around violations of these assumptions, and (iv)
has opened up a range of relevant estimation methods.

Within the POF, the regression-based {[}\emph{8}{]} and the
simulation-based {[}\emph{9}{]} approaches are widely used for the
estimation of causal mediation effects. The regression-based approach
requires fitting parametric regressions models for the mediator and the
outcome and involves approximations in the case of binary outcomes and
mediators. On the other hand, the simulation-based approach is quite
flexible and can accommodate parametric and non-parametric models. The
regression-based approach implemented in SAS and SPSS macros relies on
frequentist methods and the simulation-based approach implemented in the
widely used \textbf{mediation} R package {[}\emph{10}{]} is based on the
quasi-Bayesian approximation where the posterior distribution of
quantities of interest is approximated by their sampling distribution.

Recently, Bayesian modelling has been introduced to the mediation
analysis literature {[}\emph{11}, \emph{12}{]}. Compared to conventional
frequentist mediation analysis, the Bayesian approach combined with POF
has several advantages, including accuracy in small samples, the ability
to construct credible intervals for direct and indirect effects in a
straightforward manner, probabilistic interpretation of results, and the
option of using relevant prior information {[}\emph{11}, \emph{13}{]}.
However, the open-source software tools developed so far, such as
\textbf{bmlm {[}}\emph{14}\textbf{{]}} and \textbf{bayestestR
{[}}\emph{15}\textbf{{]}}, have mainly focused on the Bayesian
implementation of the product-method or linear structural equation
modelling (LSEM) approach {[}\emph{16}{]}. The LSEM framework has been
criticised for its limited applicability beyond specific statistical
models. In this paper, we introduce a Bayesian estimation procedure and
open-source software tool, \textbf{BayesGmed}, for causal mediation
analysis using the Bayesian g-formula approach. The proposed method
follows the potential outcomes framework for effect definition and
identification. We illustrate the applicability of the proposed method
and software tool using data from MUSICIAN trial -- a randomised
controlled study {[}\emph{17}{]}.

\hypertarget{case-study-muscian-trial}{%
\section{\texorpdfstring{Case study: MUSCIAN trial
}{Case study: MUSCIAN trial }}\label{case-study-muscian-trial}}

To illustrate the methodology presented in this paper and demonstrate
the use of the R-package \textbf{BayesGmed}, we used data from the
MUSICIAN trial (Managing Unexplained Symptoms (CWP) In Primary Care:
Involving Traditional and Accessible New Approaches (ClinicalTrials.gov
Identifier: ISRCTN67013851)).

The MUSICIAN study was a 2x2 factorial trial that estimated the clinical
effectiveness and cost-effectiveness of remotely (by telephone)
delivered cognitive-behavioral therapy (tCBT), an exercise program, and
a combined intervention of tCBT plus exercise, compared with treatment
as usual (TAU) among people with CWP. For a complete discussion about
the study and setting, we refer to {[}\emph{17}{]}. Brieﬂy, a total of
442 patients with CWP (meeting the American College of Rheumatology
criteria) were randomised to one of the four treatment arms. The primary
outcome was a 7-point patient global assessment scale of change in
health since trial enrollment (range: 1: very much worse to 7: very much
better) assessed at baseline and 6 months (intervention end) and 9
months after randomisation. A positive outcome was defined as "much
better" or "very much better." Secondary outcomes including the Tampa
Scale for Kinesiophobia (TSK) {[}\emph{18}{]}(to measure fear of
movement; score range, 17-68), the Vanderbilt Pain Management Inventory
(VPMI) {[}\emph{19}{]}(for assessing active and passive coping strategy
use), and the Sleep Scale {[}\emph{20}{]} (to measure sleep quality;
score range, 0-20; higher scores indicate more sleep disturbance) were
also assessed at baseline, 6 month and 9 months after randomisation.

Previous analysis of the MUSICIAN trial data has shown a significant
benefit of tCBT in people with chronic pain as compared to treatment as
TAU {[}\emph{17}{]}. However, little is known about the mechanisms that
lead to improvement. In this paper, using the MUSICIAN trial data, we
aim to test the hypothesis that the effect of tCBT on the primary
outcome is mediated by reductions in fear of movement, passive coping
strategies, and sleep problems and an increase in the use of active
coping strategies {[}Figure 1{]}.
\begin{center}
\includegraphics[width=1\textwidth]{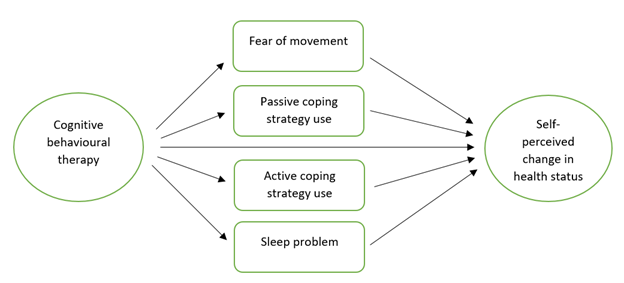}
\end{center}
Figure 1. Causal directed acyclic graph (DAG) for the MUSICIAN study.

The analysis in this paper focuses on the outcome measured six months
after randomisation and compares tCBT with treatment as usual. Baseline
characteristics of the study cohort and outcome distribution at 6 months
are presented in Table 1.

Table 1: Baseline characteristics of study cohort and outcome at 6
months post-randomisation.

\begin{longtable}[]{@{}
  >{\raggedright\arraybackslash}p{(\columnwidth - 4\tabcolsep) * \real{0.6534}}
  >{\raggedright\arraybackslash}p{(\columnwidth - 4\tabcolsep) * \real{0.1733}}
  >{\raggedright\arraybackslash}p{(\columnwidth - 4\tabcolsep) * \real{0.1733}}@{}}
\toprule()
\begin{minipage}[b]{\linewidth}\raggedright
Characteristics
\end{minipage} & \begin{minipage}[b]{\linewidth}\raggedright
TAU
\end{minipage} & \begin{minipage}[b]{\linewidth}\raggedright
tCBT
\end{minipage} \\
\midrule()
\endhead
\multicolumn{3}{@{}>{\raggedright\arraybackslash}p{(\columnwidth - 4\tabcolsep) * \real{1.0000} + 4\tabcolsep}@{}}{%
Baseline} \\
N & 109 & 112 \\
\begin{minipage}[t]{\linewidth}\raggedright
Gender

\begin{quote}
Female, n (\%)
\end{quote}
\end{minipage} & 76 (69.72) & 80 (71.42) \\
Age, mean (SD) & 56.4 (12.5) & 56.6 (13.7) \\
\multicolumn{3}{@{}>{\raggedright\arraybackslash}p{(\columnwidth - 4\tabcolsep) * \real{1.0000} + 4\tabcolsep}@{}}{%
Outcome at 6 month} \\
\begin{minipage}[t]{\linewidth}\raggedright
Perceived health status since baseline

\begin{quote}
Much better or very much better, n (\%)
\end{quote}
\end{minipage} & 7 (6.42) & 26 (23.21) \\
Fear of movement (Kinesiophobia) , mean (SD) & 36.0 (6.75) & 34.2
(6.31) \\
Active coping strategy use, mean (SD) & 24.5 (4.50) & 25.4 (4.15) \\
Passive coping strategy use, mean (SD) & 28.0 (8.13) & 27.6 (7.60) \\
Sleep problems, mean (SD) & 9.96 (6.03) & 7.83 (5.61) \\
\bottomrule()
\end{longtable}

\hypertarget{the-mathematical-framework-for-causal-mediation-analysis}{%
\section{The Mathematical Framework for Causal Mediation
Analysis}\label{the-mathematical-framework-for-causal-mediation-analysis}}

In this section, we start by reviewing the ingredients of causal
mediation analysis including definition of causal estimands/effects and
the identification assumptions needed to learn those effects from
observed data. We then describe how those causal estimands can be
estimated from observed data using the Bayesian g-formula approach. To
simplify our presentation, we restrict our examples to the context of an
observed set of time-fixed variables.

\hypertarget{definition-of-causal-mediation-effects}{%
\subsection{\texorpdfstring{Definition of Causal mediation effects
}{3.1 Definition of Causal mediation effects }}\label{definition-of-causal-mediation-effects}}

The first step in causal mediation analysis is defining the causal
effects of interest. We will start with the definition of the total
treatment effect and then introduce the direct and indirect effects.

Consider estimation of the causal effect of a binary treatment
assignment \(A\  \in \ \{ 0,\ 1\}\) on some observed outcome \(Y\),
where 1 and 0 stand for the treatment and control conditions. Following
the potential outcome framework concept {[}\emph{1}{]}, we denote the
potential outcome that would have been observed for an individual had
the exposure \(A\ \)been set to the value \(a\) by \(Y(a)\). For the
dichotomous treatment, we denote the outcome variable for the \(ith\)
individual that would have been observed under the treatment value
\(\alpha\  = \ 1\) by \(Y_{i}(1)\) and the outcome variable for the
\(ith\) individual that would have been observed under the treatment
value \(\alpha\  = \ 0\) by \(Y_{i}(0)\). Individual causal effects are
defined as a contrast of the values of these two potential outcomes and
treatment $A$ has a causal effect on an individual\textquotesingle s
outcome \(Y\) if \(Y_{i}(1) \neq Y_{i}(0)\). More formally, the total
treatment effect at the individual level is defined on additive scale as
\({TE}_{i} = Y_{i}(1) - Y_{i}(0)\). However, we never observe both
potential outcomes for the same individual. What we observe is the
realised outcome \(Y_{i}\) -- the one corresponding to the treatment
value experienced by the individual. Hence, identifying individual
causal effects is generally not possible. However, under some
assumptions to be discussed in the next subsection, the average total
effect (ATE) in a population of individuals can be estimated from the
observed data and it is defined as the average of the individual total
effects over the population. That is,
\(ATE = E\left\lbrack Y(1) - Y(0) \right\rbrack\). Put simply, the ATE
is interpreted as the average difference in the outcome had everyone in
the target population received treatment \(A\  = \ 1\) rather than
\(A\  = \ 0\). If the outcome is binary (coded 0/1), this definition is
equivalent to \(ATE\  = \ P(Y\ (1) = \ 1)\  - \ P(Y\ (0) = \ 1)\), a
risk difference. Further, given pre-exposure or pre-treatment assignment
variables \(\mathbf{Z}\), the conditional average total effect is given
by \(E\left\lbrack Y(1) - Y(0)|\mathbf{Z} \right\rbrack\).

Mediation analysis moves beyond calculation of average total treatment
effects and instead seeks to explain the effect of the exposure on the
outcome. This is achieved by splitting the total treatment effect in to
direct and indirect effects (Figure 2). By extending the previous
notations to a joint exposure \((A,\ M)\) with \(M\) being the potential
mediator, definition of direct and indirect effects can be constructed
as follows.
\begin{center}
\includegraphics[width=2.5in,height=1.2in]{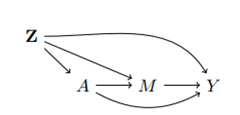}
\end{center}
Figure 2: Mediation with a single mediator M, exposure A, outcome Y, and
confounders Z.

Let \(M_{i}(a)\) denote the potential value of a mediator of interest
under the treatment status \(A = a\) and let \(Y_{i}(a,m)\) represent
the potential outcome values under regime \(A = a\) when the mediator
\(M\) is set to the value it would naturally take under either
\(A = a\). For a dichotomous exposure, the average controlled direct
effect for mediator at level m given covariate \textbf{Z} is given by
({[}\emph{1-3}{]})
\begin{equation}
CDE\ (m) = E\lbrack Y(1,m) - Y(0,m)|\mathbf{Z}\rbrack.
\end{equation}
The controlled direct effect expresses the exposure effect that would be
realised if the mediator were controlled, i.e., set to a specific level
for everyone. Controlled direct effects are relevant quantities when
interest lies in the evaluation of an intervention that can shift or fix
the mediator across the population. However, the controlled direct
effect does not usually lead to the splitting of the total effect in to
direct and indirect effect. That is, the total effect minus the
controlled direct effect may not have the interpretation of indirect
effect in situations where the direct effect is different at different
levels of the mediator. Hence, we introduce below two additional
quantities that can split the total effect in to direct and indirect
effect. They are the natural direct and natural indirect effects.

The average natural direct and indirect effects, given a pre-exposure
covariates Z, are defined as ({[}\emph{1-3}{]})

\begin{equation}
NDE(a) = E\left\lbrack Y\left( 1,M(a) \right) - Y\left( 0,M(a) \right) \middle| \mathbf{Z} \right\rbrack,
\end{equation}
and
\begin{equation}
	NIE(a) = E\lbrack Y(a,M(1)) - Y(a,M(0))|\mathbf{Z}\rbrack.
\end{equation}

The indirect effect \(NIE\) represents the causal effect of the
treatment on the outcome that can be attributed to the treatment-induced
change in the mediator and the direct effect \(NDE\ \) denotes the
causal effect of the treatment on the outcome that can be attributed to
causal mechanisms other than the one represented by the mediator, and
their sum leads to the total effect. That is,
\(TE = NIE(1) + NDE(0) = NIE(0) + NDE(1)\). Note that, \(NIE(1)\) and
\(NIE(0)\) may not be identical and a similar inequality holds for
\(NDE(1)\) and \(NDE(0)\).

\hypertarget{identification-assumptions}{%
\subsection{\texorpdfstring{Identification Assumptions
}{Identification Assumptions }}\label{identification-assumptions}}

To be able to identify or estimate the causal effects defined in 3.1, we
need to rely on a set of assumptions. To estimate the above causal
estimands from the observed data and ensure they have a causal
interpretation, the following four conditions need to hold:

\begin{itemize}
\item
  \(IA1:\ Y(a,m)\bot A|Z\): no-unmeasured confounder for the
  exposure-outcome relationship given the pre-exposure covariate Z.
\item
  \(IA2:\ Y(a,m)\bot M|A,Z\): no-unmeasured confounder for the
  mediator-outcome relationship given the pre-exposure covariate Z and
  the exposure A.
\item
  \(IA3:\ M(a)\bot A|Z\): no-unmeasured confounder for the
  exposure-mediator relationship given the pre-exposure covariate Z.
\item
  \(IA4:Y(a,m)\bot M(a^{*})|Z\) for any value of \(a,a^{*},\ and\ m\):
  no-measured or unmeasured confounder for the mediator-outcome
  relationship that is also influenced by the exposure.
\end{itemize}

Under assumptions IA1-IA4, the natural direct and indirect effects can
be identified ({[}\emph{2}, \emph{3}, \emph{5}{]}) by

NDE:
\begin{equation}
		\begin{split}
&E\left\lbrack Y\left( a,M\left( a^{'} \right) \right) - Y\left( a^{'},M\left( a^{'} \right) \right) \middle| \mathbf{Z} \right\rbrack =\\
&\int\int\left\{ E\left\lbrack Y_{i} \middle| M_{i} = m,\ A_{i} = a,\ \mathbf{Z}_{i} = \mathbf{z} \right\rbrack 
- E\left\lbrack Y_{i} \middle| M_{i} = m,\ A_{i} = a^{'},\ \mathbf{Z}_{i} = \mathbf{z} \right\rbrack\right\}\\
&\times dF_{M_{i}|A_{i} = a,\ \mathbf{Z}_{i} = \mathbf{z}}(m)dF_{Z_{i}}\left( \mathbf{z} \right),
\end{split}
\end{equation}

and

NIE:
\begin{equation}
	\begin{split}
&E\left\lbrack Y\left( a,M(a) \right) - Y\left( a,M\left( a^{'} \right) \right) \middle| \mathbf{Z} \right\rbrack =\\
&\int\int E\left\lbrack Y_{i} \middle| M_{i} = m,\ A_{i} = a,\ \mathbf{Z}_{i} = z \right\rbrack\\
&\times \left\{ dF_{M_{i}|A_{i} = a,\ \mathbf{Z}_{i} = \mathbf{z}}(m) - dF_{M_{i}|A_{i} = a^{'},Z_{i} = \mathbf{z}}(m) \right\} 
\times dF_{Z_{i}}(\mathbf{z}).
\end{split}
\end{equation}

If the mediator is discrete, the integrals will be replaced by summation
over the possible values of \(M\). In the epidemiological literature,
computation of causal effects using the above expression is called
standardisation -- a special case of g-computation.

Note that, to identify the control direct effect, only assumption IA1
and IA2 are need to hold. If assumption \(IA1\) and \(IA2\) hold, then
the controlled direct effects are identified {[}\emph{2}{]} by

\begin{equation}
	E\left\lbrack Y(1,m) - Y(0,m) \middle| \mathbf{Z} \right\rbrack = E\left\lbrack Y \middle| A = 1,M = m,\mathbf{Z} \right\rbrack - E\left\lbrack Y \middle| A = 0,M = m,\mathbf{Z} \right\rbrack,
\end{equation}

and the average controlled direct effect can be estimated from the data
by averaging over the distribution of \textbf{Z}.

\hypertarget{estimation}{%
\subsection{\texorpdfstring{Estimation
}{Estimation }}\label{estimation}}

After defining the causal estimands and specifying the necessary
conditions for the estimand to be identified, the next step is doing the
actual estimation from the observed data. In this section, we will
introduce Bayesian modeling for causal effect estimation. Bayesian
causal mediation analysis combines Bayesian modeling with the
identification assumptions discussed in 3.2 to compute a posterior
distribution over the causal estimands of interest.

Suppose we observe data
\(D\  = \ \left\{ {Y_{i},\ M_{i},\ A}_{i},\ \mathbf{Z}_{i} \right\}_{i = 1:n}\)
on \(n\ \)independent individuals, where \(A_{i} \in \ \{ 0,\ 1\}\) is a
binary treatment indicator, \(\mathbf{Z}_{i}\) is a vector of
confounders, \(M_{i}\) is a scalar candidate mediator, and \(Y_{i}\) is
a binary outcome of interest. Assume assumption \(IA1 - IA6\) hold, and
that and that the following regression models for \(Y\) and \(M\) are
correctly specified:
\begin{align}
&\mbox{logit}\left( P(Y_{i} = 1|A_{i},M_{i},\mathbf{Z}_{i}) \right) = \alpha_{0} + \mathbf{\alpha}_{Z}^{'}\mathbf{Z}_{i} + \alpha_{A}A_{i} + \alpha_{M}M_{i},\\
&E\left\lbrack M_{i} \middle| \left( A_{i},\mathbf{Z}_{i} \right) \right\rbrack = \beta_{0} + \mathbf{\beta}_{Z}^{'}\mathbf{Z}_{i} + \beta_{A}A_{i},\ \ \mbox{with}\ \epsilon_{i} \sim N(0,\ \sigma^{2}).
\end{align}

In addition to the probability model for the conditional distribution of
the outcome and the mediator (the likelihood), Bayesian inference
requires a probability distribution over the unknown parameter vector,
\(\mathbf{\theta} = (\alpha_{0},\mathbf{\alpha}_{z},\ \alpha_{A},\ \alpha_{M},\ \beta_{0},\ \mathbf{\beta}_{Z},\ \beta_{A})\),
governing this conditional distribution (i.e. a prior). Inference then
follows from making probability statements about \(\mathbf{\theta}\)
having conditioned on the observed data (via the posterior). One of the
key advantages of Bayesian inference is using priors one can obtain a
stabilised causal effect estimates when data are sparse. Specification
of priors to induce shrinkage is beyond the scope of this paper and we
refer interested readers to {[}\emph{21}{]}. For now, we assume suitable
priors in line with the specific problem one is addressing are
specified.

Bayesian estimation of causal effects rely on Bayesian analog of the
g-formula (standardisation) and bootstrap estimation of the confounder
distribution. The Bayesian analog to the g-formula {[}\emph{22}{]}
formulates the distribution of the counterfactual \(Y_{a}\) as a
posterior predictive value, integrating over the parameters
\(\mathbf{\theta}\) as well as the confounder distribution.

\[p\left( \widetilde{y}(a) \middle| o \right) = \int\int p\left( \widetilde{y} \middle| a,\widetilde{\mathbf{z}},\theta \right)p\left( \widetilde{\mathbf{z}} \middle| \mathbf{\theta} \right)p\left( \mathbf{\theta} \middle| o \right)d\mathbf{\theta}d\widetilde{\mathbf{z}}.\]

Integration over the parameters and the confounder distribution as well
as the computation of causal effects involve the following 5 steps.

\begin{enumerate}
\def\labelenumi{\arabic{enumi}.}
\item
  Given \(B\) iterations, at the \(b^{th}\) iteration obtain the
  posterior draws of the parameters \(\mathbf{\theta}\) and denote them
  by
  \(\mathbf{\theta}^{(b)} = (\alpha_{0}^{(b)},\mathbf{\alpha}_{z}^{(b)},\ \alpha_{A}^{(b)},\ \alpha_{M}^{(b)},\ \beta_{0}^{(b)},\ \mathbf{\beta}_{Z}^{(b)},\ \beta_{A}^{(b)})\).
\item
  Using the classical bootstrap, sample \(n\) new values of
  \(\mathbf{Z}\) with replacement from the observed \(\mathbf{Z}\)
  distribution during iteration \(b\) of the Markov Chain Monte Carlo
  and denote these resampled values as
  \(\mathbf{Z}^{(1,b)},\ldots,\mathbf{Z}^{(n,b)}\).
\item
  Get the potential outcome values

  \begin{enumerate}
  \def\labelenumii{\alph{enumii}.}
  \item
    Simulate the potential values of the mediator. Using the resampling
    of \(\mathbf{Z}\) as described earlier, we can draw samples from the
    distributions of the counterfactuals \(M(a)\) for
    \(a \in \{ 0,1\}\). At the \(b^{th}\) MCMC iteration and for
    \(i = 1,\ldots,n\),
  \end{enumerate}
\end{enumerate}

\[M(a)^{(i,b)} \sim Normal\left( \beta_{0}^{(b)} + \mathbf{\beta}_{Z}^{(b)}\mathbf{Z}^{(i,b)} + \beta_{A}^{(b)}a,\sigma^{(b)} \right)\]

\begin{enumerate}
\def\labelenumi{\alph{enumi}.}
\setcounter{enumi}{1}
\item
  Given the potential value for the mediator, simulate the potential
  value for the outcome. For example, \(Y(a,M(a)^{(i,b)})^{(i,b)}\) is
  simulated using
\end{enumerate}

\begin{align*}
&Y(a,M(a)^{(i,b)})^{(i,b)} \sim \\ 
& Bernoulli\left( {logit}^{- 1}\left( \beta_{0}^{(b)} + \mathbf{\beta}_{Z}^{(b)}\mathbf{Z}^{(i,b)} + \beta_{A}^{(b)}a + \alpha_{M}^{(b)}*M(a)^{(i,b)} \right) \right)
\end{align*}

\begin{enumerate}
\def\labelenumi{\arabic{enumi}.}
\setcounter{enumi}{3}
\item
  Compute draw of the causal effect estimates.

  \begin{enumerate}
  \def\labelenumii{\alph{enumii}.}
  \item
    \(ND{E(a)}^{(b)} = \frac{1}{n}\sum_{i = 1}^{n}{\{ Y\left( a',M(a')^{(i,b)} \right)^{(i,b)} - Y\left( a,M\left( a^{'} \right)^{(i,b)} \right)^{(i,b)}\}}\)
  \item
    \(NI{E(a)}^{(b)} = \frac{1}{n}\sum_{i = 1}^{n}{\{ Y\left( a,M(a)^{(i,b)} \right)^{(i,b)} - Y\left( a,M\left( a^{'} \right)^{(i,b)} \right)^{(i,b)}\}}\)
  \end{enumerate}
\item
  Get summary of causal effect estimates by taking the mean and
  quantiles of the causal effect estimates draws.
\end{enumerate}

\hypertarget{sensitivity-analysis}{%
\subsection{\texorpdfstring{Sensitivity Analysis
}{Sensitivity Analysis }}\label{sensitivity-analysis}}

As described in section 3.1, estimating direct and indirect effects from
observed data requires a series of assumptions. As a result, the main
challenge in mediation analysis has been understanding bias from
unmeasured confounding variables. Several methods have been proposed in
the literature to explore the sensitivity of causal effect estimates to
unmeasured confounding {[}\emph{23},\emph{24}, \emph{26}, \emph{27}{]}. In
our Bayesian causal mediation analysis R-package, presented in the
following section, we implemented the Bayesian sensitivity analysis
(BSA) proposed by {[}\emph{26},\emph{27}{]}. BSA works by incorporating
uncertainty about unmeasured confounding in the outcome and mediator
model through a prior distribution. That is, we extend the outcome and
mediator model in Equations (7) and (8) to a triple set of structural
equations
\begin{align}
&\mbox{logit}\left( P(Y_{i} = 1|A_{i},M_{i},\mathbf{Z}_{i}) \right) = \alpha_{0} + \mathbf{\alpha}_{Z}^{'}\mathbf{Z}_{i} + \alpha_{A}A_{i} + \alpha_{M}M_{i} + \alpha_{U}U_{i},\\
&E\left\lbrack M_{i} \middle| \left( A_{i},\mathbf{Z}_{i} \right) \right\rbrack = \beta_{0} + \mathbf{\beta}_{Z}^{'}\mathbf{Z}_{i} + \beta_{A}A_{i} + \beta_{U}U_{i},\ \ with\ \epsilon_{i} \sim N\left( 0,\ \sigma^{2} \right),\\
&\mbox{logit}\left( P(U_{i} = 1|A_{i},\mathbf{Z}_{i}) \right) = \gamma_{0} + \gamma_{A}A_{i},
\end{align}
where the binary random variable~\emph{U}~that takes values 1 or 0
indicates the presence or absence of an unmeasured confounder and the
parameters \(\alpha_{U}\ \)and \(\beta_{U}\ \)governs the association
between~\emph{U}~and Y and U and~\emph{M,} respectively. Finally\emph{,}
\(\gamma_{0}\) and \(\gamma_{A}\ \)controls the prevalence of the
unmeasured confounder within levels of~the exposure variable A~given Z.

The BSA approach proceeds by assuming a uniform prior distribution,
\(Uniform( - \delta,\ \delta),\ \)for the bias parameters
\(\alpha_{U},\ \beta_{U},\ \gamma_{0}\ \)and \(\gamma_{A}\) where
\(\delta\ \)to represent the size of unmeasured confounding (E.g.
\(\delta = 0\ \)means no unmeasured confounding){[}\emph{26},\emph{27}{]}. The
estimation of direct and indirect effect using Equations 9 -- 11
follows the same procedure as described in section 3.3 but the potential
outcome and mediator values now will also depend on the values of U.
This way, the posterior distribution for the causal effect estimates
incorporates uncertainty from bias (systematic error) in addition to
uncertainty from random sampling (random error).

\hypertarget{implementation}{%
\section{\texorpdfstring{Implementation
}{Implementation }}\label{implementation}}

The \textbf{BayesGmed} package implements Bayesian causal mediation
analysis procedure described in the previous section in R using the
probabilistic programming language \textbf{Stan} \emph{{[}28{]}}. The
latest development version of the R-package, \textbf{BayesGmed}, can be
installed from GitHub via:
\begin{quote}
\begin{verbatim}
devtools::install_github("belayb/BayesGmed”)
\end{verbatim}
\end{quote}

Models are fitted in \textbf{BayesGmed} using the following procedure:

\begin{quote}
\begin{verbatim}
bayesgmed(outcome, mediator, treat, covariates = NULL, 
          dist.y = “continuous”, dist.m = “continuous”,  
          link.y = “identity”, link.m = “identity”, data, 
          control.value = 0, treat.value = 1, 
          priors = NULL, …)
\end{verbatim}
\end{quote}

The \textbf{BayesGmed} R-package currently handles continuous outcome --
continuous mediator, binary outcome -- binary mediator, continuous
outcome -- binary mediator, and binary outcome -- continuous mediator. Currently, a multinormal, $\mbox{MVN}(\text{location}, \text{scale})$, prior is assigned to all regression parameters where the location and scale parameters are fixed to the following default values. The user can change the location and scale parameters by passing the location and scale parameters of the priors as a list as below 

\begin{quote}
	\begin{verbatim}
priors <- list(scale_m = 2.5*diag(P+1), 
               scale_y = 2.5*diag(P+2),
               location_m = rep(0, P+1), 
               location_y = rep(0, P+2),
               scale_sd_y = 2.5, 
               scale_sd_m = 2.5)
\end{verbatim}
\end{quote}
where $P$ is the number of covariates (including the intercept) in the mediator/outcome model. For the residual standard deviation, a half-normal prior is assumed with mean zero. The user can change the $scale_sd$ values as above. 

To conduct sensitivity analysis, the \emph{bayesgmed\_sens} function in
\textbf{BayesGmed} can be used as follow:

\begin{quote}
\begin{verbatim}
bayesgmed_sens(outcome, mediator, treat, covariates = NULL, 
               dist.y = “continuous”, dist.m = “continuous”,  
               link.y = “identity”, link.m = “identity”, 
               data, control.value = 0, treat.value = 1, 
               priors = NULL, …)
\end{verbatim}
\end{quote}

The \emph{bayesgmed\_sens} function have the same structure as the main
function \emph{bayesgmed} except one has to provide a list of priors for
the bias parameters. The user has then has to call \emph{bayesgmed\_sens} multiple times for a varying level prior scale parameter. Detailed vignettes describing the step-by-step use
of \textbf{BayesGmed} to conduct causal mediation analysis on various
types of outcomes and mediators are currently available at
\url{https://github.com/belayb/BayesGmed}\emph{.}

\hypertarget{results}{%
\section{Results}\label{results}}

We analysed the MUSICIAN trial data using the Bayesian causal mediation
analysis framework presented in the previous section and implemented in
the R-package \textbf{BayesGmed}. We investigated the potential
mediating effect of each of the mediators separately, assuming
independence between the mediators. We considered a logistic regression
model for the outcome and a linear regression model for the mediator
model (see Appendix S1). For all model parameters, we assumed
non-informative priors listed in Appendix S1. We ran 4 Markov chain
cycles, each with 4000 samples after 4000 burn-in samples and assessed
convergence using standard MCMC convergence checks. For a simple
comparison of the \textbf{BayesGmed} result with the result of the
well-established method, we also analysed the data using the
\textbf{mediation} R-package and presented the results side by side.

Compared to TAU, we found that tCBT has a significant positive effect on
self-perceived change in health status (Table 2). The adjusted log-odds
of tCBT on self-perceived change in health status compared to TAU range
from 1.491 (95\% CI: 0.452 -- 2.612) when adjusted for sleep problems to
2.264 (95\% CI: 1.063 - 3.610) when adjusted for fear of
movement.\textbf{~} Adjusted for the intervention, the result of the
outcome model revealed a significant relationship between self-perceived
change in health status and fear of movement, passive coping, and sleep
problem. Higher scores of fear of movement, passive coping, and sleep
problem leads to lower odds of a positive self-perceived change in
health status.~However, the result of the mediator model shows that tCBT
has a significant influence only on reducing sleep problem score
(-2.350, 95\% CI: -4.132, -0.569). tCBT had a negative relationship with
fear of movement and passive coping score and a positive relationship
with the active coping score but none of them are statistically
significant.\textbf{~}

The result of~\textbf{BayesGmed}~shows that none of the mediated effects
are statistically significant, indicating that either the effect of tCBT
on self-perceived change in health status is through other mechanisms
independent of fear of movement, the use of active or passive coping
strategies, and sleep problems or the study is too small to detect a
significant mediated effect. The result of~\textbf{BayesGmed}~is
comparable to the~\textbf{mediation}~R- package results except for the
indirect effect estimates of sleep problems. Analysis using
the~\textbf{mediation}~R-package shows a significant mediating effect of
sleep problems. This is due to the relatively larger standard errors
from~\textbf{BayesGmed}~since it accounts additional sources of
uncertainty in the parameter estimation.~

Table 2: MUSICIAN trial: Mediation analysis with one mediator at a time
approach. The Total effect, the average causal direct (ADE) and indirect
effects (ACME) are presented in risk difference scale. The coefficients
in the outcome model are in log odds scale and the coefficients of the
mediator model are on a linear scale. All models are adjusted for age,
sex and baseline GHQ median scores.

\begin{longtable}[]{@{}
  >{\raggedright\arraybackslash}p{(\columnwidth - 10\tabcolsep) * \real{0.2223}}
  >{\raggedright\arraybackslash}p{(\columnwidth - 10\tabcolsep) * \real{0.2021}}
  >{\raggedright\arraybackslash}p{(\columnwidth - 10\tabcolsep) * \real{0.1547}}
  >{\raggedright\arraybackslash}p{(\columnwidth - 10\tabcolsep) * \real{0.1341}}
  >{\raggedright\arraybackslash}p{(\columnwidth - 10\tabcolsep) * \real{0.1385}}
  >{\raggedright\arraybackslash}p{(\columnwidth - 10\tabcolsep) * \real{0.1483}}@{}}
\toprule()
\begin{minipage}[b]{\linewidth}\raggedright
\end{minipage} & \begin{minipage}[b]{\linewidth}\raggedright
\end{minipage} &
\multicolumn{4}{>{\raggedright\arraybackslash}p{(\columnwidth - 10\tabcolsep) * \real{0.5756} + 6\tabcolsep}@{}}{%
\begin{minipage}[b]{\linewidth}\raggedright
\begin{center}
	 Mediators
	 \end{center}
\end{minipage}} \\
\midrule()
\endhead
R-packages& & Fear of movement & Active coping & Passive coping & Sleep
problems \\
\hline\\
\multirow{13}{*}{\textbf{BayesGmed}} & Outcome Model & & & & \\
& \begin{minipage}[t]{\linewidth}\raggedright
\begin{quote}
tCBT
\end{quote}
\end{minipage} & \textbf{2.26 (1.06, 3.61)} & \textbf{1.18 (0.61,
3.31)} & \textbf{1.77 (0.27, 3.43)} & \textbf{1.49 (0.45,
2.61)} \\
& \begin{minipage}[t]{\linewidth}\raggedright
\begin{quote}
Mediators
\end{quote}
\end{minipage} & \textbf{-0.14 (-0.25, -0.05)} & -0.03 (-0.17,
0.12) & \textbf{-0.22 (-0.35, -0.10)} & \textbf{-0.18 (-0.29,
-0.08)} \\
\cline{2-6}\\
& Mediator Model & & & & \\
& \begin{minipage}[t]{\linewidth}\raggedright
\begin{quote}
tCBT
\end{quote}
\end{minipage} & -1.78 (-3.82, 0.37) & 0.52 (-0.98, 2.05) & -0.55
(-3.08, 1.98) & \textbf{-2.35 (-4.13, -0.57)} \\
\cline{2-6}\\
& Direct \& indirect effects & & & & \\
& ADE (control) & \textbf{0.21 (0.09, 0.35)} & \textbf{0.16 (0.04,
0.29)} & \textbf{0.11 (0.01, 0.23)} & \textbf{0.15 (0.03,
0.29)} \\
& ADE (treated) & \textbf{0.23 (0.09, 0.38)} & \textbf{0.16 (0.04,
0.29)} & \textbf{0.11 (0.01, 0.23)} & \textbf{0.18 (0.04,
0.33)} \\
& ACME (control) & 0.01 (-0.05, 0.09) & -0.00 (-0.05, 0.05) &
0.01 (-0.05, 0.07) & 0.03 (-0.04, 0.11) \\
& ACME (treated) & 0.04 (-0.06, 0.14) & -0.00 (-0.09, 0.09) &
0.01 (-0.09, 0.11) & 0.06 (-0.05, 0.17) \\
& Total effect & \textbf{0.25 (0.11, 0.39)} & \textbf{0.15 (0.03,
0.29)} & \textbf{0.12 (0.00, 0.25))} & \textbf{0.21 (0.07,
0.35)} \\
& ADE (average) & \textbf{0.22 (0.10, 0.35)} & \textbf{0.16 (0.05,
0.28)} & \textbf{0.11 (0.02, 0.22)} & \textbf{0.17 (0.04,
0.29)} \\
& ACME (average) & 0.03 (-0.04, 0.10) & -0.00 (-0.05, 0.05) &
0.01 (-0.06, 0.08) & 0.05 (-0.03, 0.12) \\
& & & & & \\
\hline\\
\multirow{7}{*}{\textbf{mediation}} & ADE
(control) & \textbf{0.20 (0.08, 0.35)} & \textbf{0.16 (0.06,
0.26)} & \textbf{0.11 (0.01, 0.23)} & \textbf{0.15 (0.05,
0.26)} \\
& ADE (treated) & \textbf{0.22 (0.09, 0.40)} & \textbf{0.16 (0.06,
0.27)} & \textbf{0.12 (0.02, 0.25)} & \textbf{0.17 (0.06,
0.29)} \\
& ACME (control) & 0.02 (-0.00, 0.05) & -0.00 (-0.01, 0.01) &
0.01 (-0.02, 0.04) & \textbf{0.03 (0.01, 0.07)} \\
& ACME (treated) & 0.04 (-0.00, 0.10) & -0.00 (-0.02, 0.02) &
0.01 (-0.03, 0.06) & \textbf{0.06 (0.01, 0.12)} \\
& Total effect & \textbf{0.24 (0.11, 0.41)} & \textbf{0.15 (0.06,
0.27)} & \textbf{0.12 (0.02, 0.27)} & \textbf{0.20 (0.09,
0.32)} \\
& ADE (average) & \textbf{0.21 (0.09, 0.38)} & \textbf{0.16 (0.06,
0.27)} & \textbf{0.11 (0.01, 0.24)} & \textbf{0.16 (0.06,
0.27)} \\
& ACME (average) & 0.03 (-0.00, 0.07) & -0.00 (-0.18, 0.09) &
0.05 (-0.65, 0.47) & \textbf{0.04 (0.01, 0.10)} \\
\bottomrule()
\end{longtable}

We applied BSA to the MUSICIAN trial data in order to explore
sensitivity of the results to bias from unmeasured confounding. We
considered three values for the bias parameter (i.e.,
\(\boldsymbol{\gamma} = (\gamma_0, \gamma_A, \beta_U, \alpha_U) \sim \mbox{MVN} (\mathbf{0}, \delta\mathbf{I}_4), \text{where}\ \delta = 0,\ 0.5,\ and\ 1)\) to denote varying level of departure from
no unmeasured confounder assumption. When $\delta = 0$, we fit a model without unmeasured confounder. The results of BSA are presented in
Figure 3. For brevity, we only presented the results of the average
direct (ADE), average indirect effect (ACME) and total effect (TE).
Overall, BSA leads to a much wider credible intervals for all effects of
interest than the Naive (\(\delta = 0\)). If we consider 95\% credible
interval overlap with zero in order to identify non-zero natural direct
and indirect effects,
then~\href{https://journals.sagepub.com/doi/full/10.1177/0962280217729844?casa_token=g85DXOSUmQcAAAAA\%3AouzPAdIQiYiNCbZkbWB2TBb98PIwdkCiRLz2b5SYtgK9QOTdRIw_FXfSv2-SGhTD54UoZhR07uWVkg}{Figure}
3~shows that the direct and total effect of cognitive behavioral therapy
on changes in perceived health status persists even for a large
departure in the assumption of no unmeasured confounding.
\begin{center}
\includegraphics[width=1\textwidth]{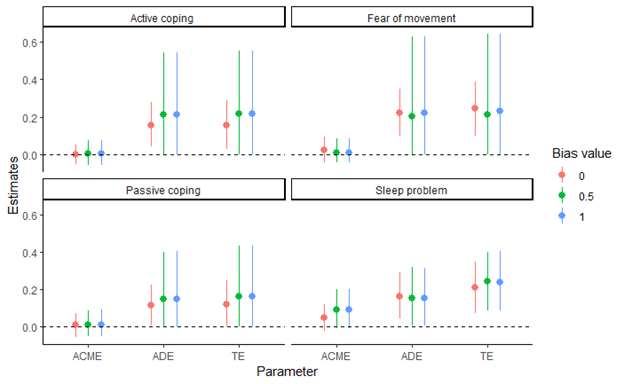}
\end{center}
Figure 3: MUSICIAN trial: Bayesian sensitivity analysis for varying
levels of departure from no-unmeasured confounder assumptions.

\hypertarget{concluding-remark}{%
\section{\texorpdfstring{Concluding remark
}{Concluding remark }}\label{concluding-remark}}

In this paper, we introduced a Bayesian estimation algorithm for causal
mediation analysis. We also provide an easy-to-use R-package for
conducting Bayesian causal mediation analysis and assessing sensitivity
of results for unmeasured confounder. Compared to the existing
open-source tools for mediation analysis, \textbf{BayesGmed} has several
advantages. First, point and interval estimates can be easily
constructed for causal risk ratios, odds ratios, and risk differences by
post-processing posterior draws from the fitted model. Second, priors
can be specified to obtain more stabilised causal effect estimates than
the frequentist procedure. Third, priors can also be used to conduct
probabilistic sensitivity analyses around violations of key causal
identification assumptions.

Using the proposed methodology, we analysed data from a randomised
control trial with the aim of identifying mediators of tCBT on
self-perceived change in health status in patients with chronic
widespread pain. We showed the beneficial effect of tCBT compared to
TAU, similar to previous reports {[}\emph{17}{]}. However, none of the
considered potential mediators (i.e. reduction in fear of movement,
reduction in passive coping, reduction in sleep problem, and an increase
in activing coping) were found to mediate the effect of tCBT. Except
active coping, all of the potential mediating factors were found to have
a statistically significant effect on the outcome of interest, but tCBT
had a significant effect only on reducing sleep problems leading to a
non-significant indirect effect. These results suggest that either
improving the scope of tCBT or combining tCBT with other interventions
that can target fear of movement, passive coping, and sleep problem
would increase patient benefit. However, it is important to note that
the MUSICIAN trial was not powered to detect mediators of the effect of
tCBT on outcome. tCBT was associated with change in scores for fear of
movement, active coping, passive coping, and sleep problems in the
expected direction, and the magnitude of effect was greatest for sleep
problems. Whether these would mediate the effect of tCBT in an
adeqaultely powered trial remains unknown. However, the methds presented
here would be able to address that question in an well-powered study. It
also remains possible that tCBT exerts its influence through some other
mechanism(s). It would be of interest to explore non-specific effects in
non-blinded trials sucha s MUSICIAN.

At present, there are some limitations of the package
\textbf{BayesGmed}. First of all, we assumed a parametric specification
for the outcome and mediator model. In some situations, parametric
models might be restrictive and a general non-parametric models might be
preferred. Second, we only considered the case of single mediator and
assumed no exposure mediator interaction. The Bayesian estimation
algorithm we presented is quite generic and can easily be extended to
accommodate the aforementioned limitation and we aim to extend the
\textbf{BayesGmed} package to handle the above settings in a future
version. Since the package is distributed as an open source software
users can also update the package for their own needs.

		\hypertarget{Competing interests}{%
		\section*{\texorpdfstring{Competing interests
			}{Competing interests}}\label{Competing interests}}
		None
		\hypertarget{Funding}{%
			\section*{\texorpdfstring{Funding
				}{Funding}}\label{Funding}}
		The authors received no specific funding for this work.
\hypertarget{acknowledgement}{%
\section*{\texorpdfstring{Acknowledgement
}{Acknowledgement }}\label{acknowledgement}}

The work is supported by the Centre for Epidemiology Versus Arthritis (grant
number 21755). The MUSICIAN trail, used as a case study in this paper,
is funded by versus Arthritis (grant number 20748).

\hypertarget{references}{%
\section{References}\label{references}}

1. Robins, J.M. and S. Greenland, \emph{Identifiability and
exchangeability for direct and indirect effects.} Epidemiology, 1992: p.
143-155.

2. Pearl, J., \emph{Causal inference in statistics: a gentle
introduction.} Computing science and statistics, proceedings of
Interface\textquotesingle01, 2001. \textbf{33}.

3. VanderWeele, T.J. and S. Vansteelandt, \emph{Conceptual issues
concerning mediation, interventions and composition.} Statistics and its
Interface, 2009. \textbf{2}(4): p. 457-468.

4. Imai, K., L. Keele, and T. Yamamoto, \emph{Identification, inference
and sensitivity analysis for causal mediation effects.} Statistical
science, 2010: p. 51-71.

5. Imai, K. and T. Yamamoto, \emph{Identification and sensitivity
analysis for multiple causal mechanisms: Revisiting evidence from
framing experiments.} Political Analysis, 2013: p. 141-171.

6. VanderWeele, T. and S. Vansteelandt, \emph{Mediation analysis with
multiple mediators.} Epidemiologic methods, 2014. \textbf{2}(1): p.
95-115.

7. Daniel, R.M., et al., \emph{Causal mediation analysis with multiple
mediators.} Biometrics, 2015. \textbf{71}(1): p. 1-14.

8. Valeri, L. and T.J. VanderWeele, \emph{Mediation analysis allowing
for exposure--mediator interactions and causal interpretation:
theoretical assumptions and implementation with SAS and SPSS macros.}
Psychological methods, 2013. \textbf{18}(2): p. 137.

9. Imai, K., L. Keele, and D. Tingley, \emph{A general approach to
causal mediation analysis.} Psychological methods, 2010. \textbf{15}(4):
p. 309.

10. Tingley, D., et al., \emph{Mediation: R package for causal mediation
analysis.} Journal of Statistical Software, 2014.

11. Miočević, M., et al., \emph{A tutorial in Bayesian potential
outcomes mediation analysis.} Structural equation modeling: a
multidisciplinary journal, 2018. \textbf{25}(1): p. 121-136.

12. Kim, C., et al., \emph{Bayesian methods for multiple mediators:
Relating principal stratification and causal mediation in the analysis
of power plant emission controls.} The annals of applied statistics,
2019. \textbf{13}(3): p. 1927.

13. Yuan, Y. and D.P. MacKinnon, \emph{Bayesian mediation analysis.}
Psychological methods, 2009. \textbf{14}(4): p. 301.

14. Vuorre, M. and N. Bolger, \emph{Within-subject mediation analysis
for experimental data in cognitive psychology and neuroscience.}
Behavior Research Methods, 2018. \textbf{50}(5): p. 2125-2143.

15. Makowski, D., M.S. Ben-Shachar, and D. Lüdecke, \emph{bayestestR:
Describing effects and their uncertainty, existence and significance
within the Bayesian framework.} Journal of Open Source Software, 2019.
\textbf{4}(40): p. 1541.

16. Baron, R.M. and D.A. Kenny, \emph{The moderator--mediator variable
distinction in social psychological research: Conceptual, strategic, and
statistical considerations.} Journal of personality and social
psychology, 1986. \textbf{51}(6): p. 1173.

17. McBeth, J., et al., \emph{Cognitive behavior therapy, exercise, or
both for treating chronic widespread pain.} Archives of internal
medicine, 2012. \textbf{172}(1): p. 48-57.

18. Roelofs, J., et al., \emph{The Tampa Scale for Kinesiophobia:
further examination of psychometric properties in patients with chronic
low back pain and fibromyalgia.} European Journal of Pain, 2004.
\textbf{8}(5): p. 495-502.

19. Brown, G.K. and P.M. Nicassio, \emph{Development of a questionnaire
for the assessment of active and passive coping strategies in chronic
pain patients.} Pain®, 1987. \textbf{31}(1): p. 53-64.

20. Jenkins, C.D., et al., \emph{A scale for the estimation of sleep
problems in clinical research.} Journal of clinical epidemiology, 1988.
\textbf{41}(4): p. 313-321.

21. Oganisian, A. and J.A. Roy, \emph{A practical introduction to
Bayesian estimation of causal effects: Parametric and non-parametric
approaches.} Statistics in Medicine, 2021. \textbf{40}(2): p. 518-551.

22. Keil, A.P., et al., \emph{A Bayesian approach to the g-formula.}
Statistical methods in medical research, 2018. \textbf{27}(10): p.
3183-3204.

23. VanderWeele, T.J., \emph{Bias formulas for sensitivity analysis for
direct and indirect effects.} Epidemiology (Cambridge, Mass.), 2010.
\textbf{21}(4): p. 540.

24. Daniels, M.J., et al., \emph{Bayesian inference for the causal
effect of mediation.} Biometrics, 2012. \textbf{68}(4): p. 1028-1036.

25. Imai, K. and T. Yamamoto, \emph{Identification and sensitivity
analysis for multiple causal mechanisms: Revisiting evidence from
framing experiments.} Political Analysis, 2013. \textbf{21}(2): p.
141-171.

26. Comment, Leah, Brent A. Coull, Corwin Zigler, and Linda Valeri. \emph{Bayesian data fusion for unmeasured confounding.} arXiv preprint arXiv:1902.10613 (2019).

27. McCandless, L.C. and J.M. Somers, \emph{Bayesian sensitivity
analysis for unmeasured confounding in causal mediation analysis.}
Statistical Methods in Medical Research, 2019. \textbf{28}(2): p.
515-531.

28. Carpenter, B., et al., \emph{Stan: A probabilistic programming
language.} Journal of statistical software, 2017. \textbf{76}(1).

\hypertarget{Appendix}{%
	\section*{Appendix}\label{Appendix}}

\hypertarget{Appendix S1: Model formulation for the MUSICIAN study}{%
	\subsection*{Appendix S1: Model formulation for the MUSICIAN study}\label{Appendix S1}}
Let $Y_{i}$ represent the binary response variable (i.e., $Y_i = 1$ denoting a much or very much better outcome since randomization) of the $i$th subject, $A$ denotes the treatment ($A = 1$ for tCBT and $A = 0$ for TAU), $M_{ki}$ represent the $k$th mediator variable for the $i$th subject, and $\mathbf{Z}_i$ denotes the baseline covariates to adjust for for the $i$ subject. The outcome and mediator model of the mediation analysis is then formulated as follows
\begin{align}
	&\mbox{logit}\left( P(Y_{i} = 1|A_{i},M_{i},\mathbf{Z}_{i}) \right) = \alpha_{0} + \mathbf{\alpha}_{Z}^{'}\mathbf{Z}_{i} + \alpha_{A}A_{i} + \alpha_{M_k}M_{ki},\\
	&E\left\lbrack M_{ki} \middle| \left( A_{i},\mathbf{Z}_{i} \right) \right\rbrack = \beta_{0} + \mathbf{\beta}_{Z}^{'}\mathbf{Z}_{i} + \beta_{A}A_{i},\ \ \mbox{with}\ \epsilon_{i} \sim N(0,\ \sigma_m^{2}).
\end{align}
The covariate vector $\mathbf{Z}_i$ includes age, gender, and baseline median GHQ score. We considered $K = 4$ mediators including $M_1$ - tsk (fear of movement measure), $M_2$ - active coping, $M_3$ - passive coping, and $M_4$ - sleep problems.  We fit the above structural model for each mediators separately assuming the following priors. 
\begin{align}
	&\mathbf{\alpha} = (\alpha_0, \alpha_Z, \alpha_A, \alpha_M)' \sim \mbox{MVN}(location_y, scale_y),\\
		&\mathbf{\beta} = (\beta_0, \beta_Z, \beta_M)' \sim \mbox{MVN}(location_m, scale_m),\\
		&\sigma_m^2 \sim (0, scale\_sd\_m).
\end{align}
We assume $\mathbf{0}_6$ and $\mathbf{0}_5$ for the $location_y$ and $location_m$, respectively. For the $scale_y$ and $scale_m$, we considered $10*\mathbf{I}_6$ and $10*\mathbf{I}_5$, respectively. Finally, we set the $scale_sd_m = 2.5$ for all models. 

\hypertarget{Appendix S2: R - code for the MUSICIAN study}{%
	\subsection*{Appendix S2: R - code for the MUSICIAN study}\label{Appendix S1}}

Install and load the \textbf{BayesGmed}, \textbf{Rstan}, and \textbf{mediation} packages. 
\begin{quote}
	\begin{verbatim}
		install.packages("rstan")
		devtools::install_github("belayb/BayesGmed”)
		install.packages("mediation")
		library(rstan)
		library(BayesGmed)
		library(mediation)
	\end{verbatim}
\end{quote}

\textbf{Mediation model fitting using the \textbf{BayesGmed} package with tsk (fear of movement) as a mediator. }
\begin{quote}
	\begin{verbatim} 
		
fit <- bayesgmed(outcome = "outcome", mediator = "tsk", 
                 treat = "TrT", 
                 covariates = c("gender", "age","ghqmedian"),
                 dist.y = "binary", dist.m = "continuous",
                 link.y = "logit", link.m = "identity", 
                 data = med_data, 
                 priors = list(scale_m = 10*diag(5), 
                               scale_y = 10*diag(6),
                               location_m = rep(0, 5),
                               location_y = rep(0, 6)),
                               iter=8000)
bayesgmed_summary(fit)
		\end{verbatim}
\end{quote}	
\textbf{Mediation model fitting using the \textbf{mediation} package with tsk (fear of movement) as a mediator. }
\begin{quote}
	\begin{verbatim} 
outcome_model <- glm(outcome ~ gender + age + ghqmedian +
                               Trt + tsk, 
                     data = med_data, 
                     family = binomial(link="logit"))
                     
mediator_model_tsk <- lm(tsk ~  gender + age + ghqmedian + Trt, 
                         data = med_data)
                         
                         
med.out_tsk <- mediate(mediator_model_tsk, outcome_model_tsk,
                       treat = "Trt", mediator = "tsk",
                       robustSE = TRUE, sims = 100)
summary(med.out_tsk)
                         
	\end{verbatim}
\end{quote}
We repeat the above steps for each mediators. 

\textbf{Sensitivity analysis for unmeasured confounding using the \textbf{BayesGmed} package with tsk (fear of movement) as a mediator. }
\begin{quote}
	\begin{verbatim} 
		
fit <- bayesgmed_sens(outcome = "outcome", mediator = "tsk",
                      treat = "TrT", 
		              covariates = c("gender", "age","ghqmedian"),
		              dist.y = "binary", dist.m = "continuous",
		              link.y = "logit", link.m = "identity", 
		              data = med_data, 
		              priors = list(scale_m = 10*diag(5), 
		                            scale_y = 10*diag(6),
		                            location_m = rep(0, 5),
		                            location_y = rep(0, 6),
		                         location_gamma = rep(0,4),
		                         scale_gamma = 0.5*diag(4)),
		              iter=8000)
	\end{verbatim}
\end{quote}	

\end{document}